\documentclass[a4paper, 11pt]{article}

\usepackage[utf8]{inputenc}
\usepackage[T1]{fontenc}

\usepackage{amssymb}        
\usepackage{latexsym}
\usepackage{graphicx}

\author{}

\date{}

\title{Faster and simpler approximation of stable matchings}

\author{Katarzyna Paluch \thanks{Supported by MNiSW grant number N N206 1723 33, 2007-2010.}\\
Institute of Computer Science, Wroc{\l}aw University }

\newcommand{\dowod}{\noindent{\bf Proof.~}}
\newcommand{\koniec}{\hfill $\Box$\\[.1ex]}

\newtheorem{fact}{Fact}

\newtheorem{lemma}{Lemma}
\newtheorem{theorem}{Theorem}

\begin{document}

        \maketitle
\thispagestyle{empty}

\begin{abstract}
We give a $\frac{3}{2}$-approximation algorithm for stable matchings that runs in $O(m)$ time.
The previously  best known algorithm by McDermid has the same approximation ratio but runs in $O(n^{3/2}m)$ time, where $n$ denotes the number of people and $m$ 
is the total length of the preference lists
in a given instance.
Also the algorithm and the analysis are much simpler.
We sketch the extension of the algorithm for computing stable many-to-many matchings.
\end{abstract}

\section{Introduction}

In the paper we consider a variant of the problem called {\bf Stable Matchings}, known also in the literature as {\bf Stable Marriage}.
The problem is defined as follows. We are given two sets $W$ and $U$ of women and men and each woman $w$ of $W$ has a linearly ordered preference list $L_w$ of a subset of men $U'_{w} \subseteq U$
and similarly each man $m$ of $U$ has a linearly ordered preference list $L_m$ of a subset of women $W'_{m} \subseteq W$. The lists of men and women can contain {\bf ties}, which are subsets of  men (or respectively women),
which are equally good for a given woman (resp. man). Thus if $m$ and $m'$ are on list $L_w$ of woman $w$, then either (1) $m <_w m'$ and then
we say that woman $w$ prefers $m$ to $m'$ or that  $m$ is better for her than $m'$ or (2) $m\equiv_w m'$, which means that $m$ and $m'$ are in a tie on $L_w$ and then we say that $w$ is indifferent between $m$ and $m'$ or that
$m$ and $m'$ are equally good for her or (3) $m'<_w m$. If man $m$ does not belong to list $L_w$, then we say that $m$ is {\bf unacceptable} for $w$.
A {\bf matching} is a set of pairs $(m, w)$ such that $m \in U, w \in W$ and $m$ and $w$ are on each other's preference lists and each man/woman belongs to at most one pair. 
If $(m_1,w_1)$ belongs to a certain matching $M_1$, then we write $M_1(m_1)=w_1$, which means that in $M_1$ woman $w_1$ is a partner of $m_1$ and analogously that   $M_1(w_1)=m_1$. If man $m$ (or woman $w$) is not contained in any pair of
matching $M$, then we say that $m$ ($w$) is {\bf unmatched} or {\bf single} or {\bf free} in $M$.
A matching $M_1$ is called {\bf stable} if it does not admit a {\bf blocking pair}. A pair $(m,w)$ is blocking for $M_1$  if (0) $m$ and $w$ are acceptable to each other and (1) $m$ is single or prefers $w$ to $M_1(m)$ and if (2) $w$ is single or prefers $m$ to $M_1(w)$.
Each instance of the problem can be represented by a bipartite graph $G=(U\cup W,E)$ with vertices $U$ representing men and verices $W$ women and edges $E$ connecting all mutually acceptable pairs of men and women.
The problem we are interested in in this paper is that of finding a stable matching that has the largest cardinality.
The version in which there are no ties in the preference lists  of men and women has been long known and an algorithm by Gale and Shapley (\cite{GS}) solves it exactly in $O(m)$ time, where $m$ denotes the sum of the lengths of preference lists. In the version without ties a stable matching always exists and every stable matching has the same cardinality. 
If we allow ties, as in the problem we consider in this paper, then
a stable matching also always exists and can be found via the Gale/Shapley algorithm  by breaking ties arbitrarily.    However, the sizes of stable matchings can vary considerably and
the problem of finding a stable matching of maximum cardinality is $NP$-hard, which was 
shown by Manlove et al. in \cite{Manlove}.
Therefore it is desirable to devise an approximation algorithm for the problem. 

\noindent{\bf Previous results}
Previous approximation algorithm were presented in
\cite{Manlove},\cite{Iwama1}, \cite{Iwama2}, \cite{Iwama3}, \cite{Kir}. Currently the best approximation algorithm is by  McDermid \cite{McDermid} and achieves the approximation guarantee $\frac{3}{2}$. Its  running time is $O(n^{3/2}m)$, where $n$ denotes the number of vertices and $m$ the number of edges. 
Inapproximabilty results were shown in \cite{Halld1}, \cite{Halld2}, \cite{Y}.

\noindent {\bf Our results} We give a $3/2$-approximation algorithm that runs in $O(m)$ time and additionally is significantly simpler than that of McDermid. We sketch the extension of the algorithm for computing stable many-to-many matchings, which runs in $O(m log c)$ time,
where $c$ denotes the  minimum of the  maximal capacities in each side of the bipartition.
In particular it means we give an $O(m)$-time algorithm for the Hospitals-Residents problem, improving on an   
$O(d^{5/2}n^{3/2}m)$ time algorithm given by McDermid, where $d$ denotes the maximal capacity of a hospital.
Since the problems are said to have many practical applications, we believe our algorithms will be of help.

\section{Algorithm}
Let $M_{opt}$ denote an optimal (i.e. largest) stable matching and $M, M'$ any  two matchings.  
We say that $e$ is an $M$-edge if $e \in M$.
A path $P$ or a cycle $C$ is called {\bf alternating (with respect to $M$)} if its edges alternate between $M$-edges
and edges of  $E \setminus M$. 
It is well known from matching theory (see \cite{Lov} for example) that $M \oplus M'$ can be partitioned into a set of alternating paths and alternating cycles. (For two sets $X,Y$, the set $X \oplus Y$ denotes $X \setminus Y \cup Y \setminus X$.) Let $S$ denote a set of alternating paths and cycles of $M \oplus M_{opt}$.
Consider any alternating cycle $c$ of $S$ or any alternating path $p$ of even length of $S$. Then both $c$ and $p$ contain the same number of $M$-edges and $M_{opt}$-edges. Consider an alternating path $p$ of length $2k+1$ of $S$.
Then either $\frac{|M_{opt} \cap p|}{|M \cap p|}= \frac{k+1}{k}$ or $\frac{|M \cap p|}{|M_{opt} \cap p|}= \frac{k+1}{k}$.
Therefore if $M \oplus M_{opt}$ does not contain paths of length $3$ with the middle edge of a path being an $M$-edge, then $|M_{opt}| \leq \frac{3}{2} |M|$ and $M$ is a $\frac{3}{2}$ approximation of $M_{opt}$.  
To achieve a $\frac{3}{2}$-approximation we will be eliminating such potential paths of length $3$ of $M \oplus M_{opt}$.

Accordingly we define a {\bf dangerous  path}, which  is an alternating path $P= (w,m_1,w_1,m)$ such that $w$ and $m$ are unmatched (which means that $(m_1,w_1)$ is in $M$ and $(w,m_1), (w_1,m)$ do not belong to $M$)
and $(m_1,w_1)$ is not a blocking pair for matching $M'=\{(w,m_1), (w_1,m)\}$. Let us notice that if $P$ is a dangerous  path, then either $m_1$ is indifferent between $w$ and $w_1$ (and then we say that $(w,m_1)$
is an {\bf equal edge}) or $w_1$ is indifferent between
$m$ and $m_1$ (and then $(w_1,m)$ is called an equal edge) or both. Thus a dangerous path contains one or two equal edges. If an edge $e=(w,m_1)$ is equal, then we say that $P$ is a {\bf masculine dangerous path} and if $(w_1,m)$ is equal, then we say that $P$ is a {\bf feminine dangerous path}. A path $P$ can of course be both a masculine and feminine dangerous  path.

We also introduce the following terminology.
If man $m$ is matched to woman $w$ and there is at least one free woman $w_1$ such that $w$ and $w_1$ are equally good for $m$, then we say that $m$ is {\bf unstable} and that $w_1$ is a {\bf satellite} of $m$. 
If woman $w$ is matched to an unstable man $m$, then we say that $w$ is {\bf unstable}.
If $e=(m,w)$ is such that $w$ is free and $m$ is either free or matched to $w'$, who is equally good for him as $w$ and there is at least one free woman $w_1$ such that $w$ and $w_1$ are equally good for $m$, then $e$ is called {\bf special}. 
An edge $(m,w)$ is said to be {\bf blocking} if $m,w$ are a blocking pair.
An edge $(m,w)$ is said to be {\bf unstable} if $m$ is free and $w$ is unstable.
Let us notice that if $(m,w)$ is unstable, then it is a  part of a masculine dangerous path.
An edge $(m,w)$ that is equal and such that $m$ is free is called {\bf f-equal}. 
An edge $e$ is said to be {\bf bad} if it is blocking or unstable or f-equal.

In the algorithm given below set $X$ contains single men that have not yet proposed to all women on their lists
(and potentially belong to blocking pairs or masculine dangerous paths) and set $F$ contains single men, who
have already proposed to all women on their lists and potentially belong to feminine dangerous paths.  

\noindent \fbox{
\begin{minipage}[t]{\textwidth}
\scriptsize
Algorithm {\em GS Modified}\\
\\
Each man $m$'s preference list $L_m$ is organized in such a way that if $L_m$ contains a tie $t=(w_1,w_2, \ldots, w_k)$, then free women in $t$ come before matched women in $t$.
At the beginning all women are free  and ties are broken arbitrarily and  in the course of running the algorithm whenever woman $w$ becomes matched for the first time we move her to the end of every tie she belongs to.
\\
\\
$X:=U$ (all men)\\
$F:=\emptyset $ \\

\noindent while $X$ or $F$ is nonempty \\
 if there is a man $m \in X$ , then 
 
 \hspace{1cm} if $L_m=\emptyset $, then  remove $m$ from $X$ and if $L'_m \neq \emptyset $, add $m$ to $F$ 
 
 \hspace{1cm} else 
 
 \hspace{2cm} $w \leftarrow $ next woman on $m$' s list $L_m$ 
 
 \hspace{2cm} if $(m,w)$ is not special, then remove $w$ from $L_m$ 
 
 \hspace{2cm} if $(m,w)$ is blocking, then 
 
 \hspace{5cm} add $M(w)$ to $X$
 
 \hspace{5cm} $M \leftarrow M \cup (m,w) \setminus (w, M(w))$
 
 \hspace{5cm} remove $m$ from $X$
 
 \hspace{2cm} else if $w$ is unstable, then 
 
 \hspace{5cm} let $w'$ be a satellite of $M(w)$
 
 \hspace{5cm} if $(M(w),w')$ is special, remove $w'$ from $L_{M(w)}$  
 
 \hspace{5cm} $M \leftarrow M \cup \{(m,w),(M(w),w')\} \setminus (w, M(w))$
 
 \hspace{5cm} remove $m$ from $X$
 
 \hspace{2cm} else if $(m,w)$ is f-equal, then add $w$ to the end of list $L'_m$.

 else 
 
 \hspace{1cm} there is a man $m \in F$
 
 \hspace{1cm} if $L'_m =\emptyset$, remove $m$ from $F$ 
 
 \hspace{1cm} else 
 
 \hspace{2cm} $w \leftarrow $ next woman on $m$' s list $L'_m$ 
 
 \hspace{2cm} remove $w$ from $L'_m$ 
 
 \hspace{2cm} if $(m,w)$ is f-equal, then 
 
 \hspace{5cm} add $M(w)$ to $X$
 
 \hspace{5cm} $M \leftarrow M \cup (m,w) \setminus (w, M(w))$ 
 
\end{minipage} 
}
\\

First we show how Algorithm  GS Modified runs on the following example. Suppose the preference lists of men $m_1,m_2,m_3$ and women $w_1,w_2,w_3$ are as follows. The brackets indicate ties. \\
 
$\begin{array}{lr}
\begin{array}{llll}
m_1: & (w_1, w_2) & w_3 & \\
m_2: & w_1 &w_3& w_2 \\
m_3: & w_2 &w_1 &w_3 
\end{array}
&
\begin{array}{llll}
w_1: & m_1 &m_2 & m_3\\
w_2: & m_3 & m_1 & m_2 \\
w_3: & m_1 & m_2 & m_3 
\end{array}
\end{array}$
\\

Suppose that $m_1$ starts.
$m_1$ proposes to $w_1$ and gets accepted ($(m_1,w_1)$ is a special edge and $w_2$ is a satellite of $m_1$).
Now suppose that it is $m_2$'s turn to propose. (It might also be $m_3$.)
$m_2$ proposes to $w_1$ and gets accepted because $w_1$ is unstable. $m_1$ gets matched with $w_2$.
$m_3$ proposes to $w_2$ and gets accepted. $m_1$ proposes to $w_1$ (as $(m_1,w_1)$ was a special edge) and gets accepted. $m_2$ proposes to $w_3$ and gets accepted. 

If we break ties arbitrarily and run the Gale/Shapley algorithm, then the cardinality of the computed matching depends
on the order in which we break ties and the order in which men propose. Algorithm GS Modified outputs a matching that
would have been output by the GS algorithm if the order of ties and proposals of men were as follows. It would be identical to that in Algorithm GS Modified but for two things: (1) if $m$ proposes to $w$, $(m,w)$ is not blocking,
$w$ is unstable, $w'$ is a satellite of $M(w)$, then a tie in the list of $L_{M(w)}$ would be broken so that $w'$ would come before $w$ (whereas in Algorithm GS Modified $w$ comes before $w'$) and thus $M(w)$ would propose first to $w'$
and not $w$, (2) if $m \in F$ and $(m,w)$ is f-equal, then $m$ would propose to $w$ before $M(w)$.

Now we prove the correctness of Algorithm GS Modified.

\begin{fact} \label{F}
If woman $w$ becomes matched, then she will stay matched.  Woman $w$ can become unstable only the first time someone proposes to her.
If an unstable woman $w$ (matched to some unstable man $m$) becomes matched to some new man $m'$, then she is not unstable any more.   
If woman $w$ is matched to man $m$ and is not unstable, then she will always be matched to someone at least as good for her as $m$.
\end{fact} 

\dowod If $w$ is matched and $m$ proposes to her, then there is no free woman $w'$ who is equally good for $m$ as $w$ (because then $m$ would propose to $w'$ before proposing to $w$). 
As a result if $w$ becomes matched to $m$ she will not become unstable and she will cease to be unstable if she was before. 
Also if $w$ is matched and not unstable at the moment $m$ proposes to her, she accepts him only if $m$ is not worse for her than her current partner $M(w)$. \koniec

\begin{lemma}\label{E}
If $e$ is special at some step of  Algorithm  GS Modified  and gets added to $M$, then it will not become special later.
Suppose that at  some step $S$ of Algorithm  GS Modified edge $e$ is incident with a single man $m$. If at step $S$ edge $e$  is not bad, then it will not become bad later.
If at step $S$ edge $e$ is f-equal and  not unstable, then $e$  will not become unstable or blocking.
If $e$ is bad but not special at step $S$ and gets added to $M$, then  it will not
become bad later in the course of running the algorithm.
\end{lemma}

\dowod If at step $S$ edge $e=(m,w)$ is special and gets added to $M$, then it means that  $w$ becomes matched and by Fact \ref{F} she will always stay matched, therefore $e$ will never become special. 

If $e=(m,w)$ is not bad, then $w$ is matched to some man $m'$.
 Since $e$ is not unstable, then  $w$ is  not unstable and thus by Fact \ref{F} $w$ will not become unstable, thus $e$ will not become unstable.
If $w$ prefers $m'$ to $m$, since she is not unstable, she will always be matched to someone she prefers to $m$, thus $e$ will not become blocking.
If for $w$ men $m$ and $m'$ are equally good, but $e$ is not an equal edge, then it means that $m'$ has no free woman incident on him at the moment and thus will never have
and if $w$ gets matched to $m''$, then $m''$ will be better for her than $m'$, because $w$ could become matched to $m''$ that is equally good for her as $m'$ only if $(w,m')$ belonged to a feminine dangerous 
path. 

If at step $S$ edge $e=(m,w)$ is f-equal and not unstable, then it means that $w$ is matched to some man $m'$ such that $m$ and $m'$ are equally good for her and $w$ is not unstable.
Thus by Fact \ref{F} she will not become unstable and she will always be matched to someone at least as  good for her as $m'$.

Suppose that at step $S$  edge $e=(m,w)$ is bad but not special and $e$ gets added to $M$.
Then by Fact \ref{F} $w$ will never become unstable (note that at step $S$ she might have been unstable) and will always stay matched to $m$ or will become matched to someone at least as good for her as $m$.
Thus $e$ will not become unstable or blocking later.
If she gets matched to $m'$, who is equally good for her as $m$, then it means that at that step, $(m,w)$ belonged to a femine dangerous path and thus $m'$ had not (and thus will never have) a free woman incident on him (because then he would have proposed to her as he would have been blocking because of a free woman incident on him), which means that $(w,m')$
never becomes a part of a feminine dangerous path, therefore $e$ will not become f-equal. \koniec

\begin{theorem}
Algorithm GS Modified computes
computes a stable matching $M$ that does not contain dangerous alternating paths and thus is a $\frac{3}{2}$- approximation algorithm. Algorithm  GS Modified  runs in $O(m)$ time.
\end{theorem}
\dowod Suppose that matching $M$ computed by the algorithm is not stable. Then it contains a blocking edge $(m,w)$.
Thus $m$ is either single or matched to woman $w'$, who is worse for him than $w$. Therefore at some step of the algorithm $m$ must have proposed to $w$. If at that step $(m,w)$ was blocking, it got added to $M$
and by Lemma \ref{E}, $(m,w)$ could not become blocking later and if it was not blocking, it also could not become
blocking later.

For $\frac{3}{2}$-approximation, it suffices to show that the graph does not contain dangerous paths.
If the graph contains a masculine or feminine dangerous path, then it contains an edge $e=(m,w)$, that is unstable
or f-equal. But then at some step of the algorithm $m$ proposed to $w$ and if $e$ was not unstable or f-equal then,
by Lemma \ref{E} it could not become unstable or f-equal later. If it was unstable at that step it got added to $M$
and also could not become unstable or f-equal later. If it was f-equal, $w$ was added to $L'_m$ and $e$ was considered again at some later step and either it got added to $M$ because it still was f-equal or not because it was not
and by Lemma \ref{E} could not become f-equal later.

The running time of the algorithm is proportional to the total length of lists $L_m$ and $L'_m$.
Each edge of $L_m$ is scanned at most twice - twice, only if the first time it was scanned, it was special
and each edge of $L'_m$ is scanned at most once. \koniec

\section{Extension to stable $b$-matchings}
Suppose we have a bipartite graph $G= (V, E)$, where $V= U \cup W$ and $U,W$ are disjoint sets, and a function $b: V \rightarrow N$. Then a subset $M \subseteq E$ is called a $b$-matching if for each $v \in V$ it is $deg_M(v) \leq b(v)$, where $deg_M(v)$ denotes the degree of 
vertex $v$ in a graph  $G_M= (U \cup W, M)$.  We will call vertices of $U$ - $U$-agents and vertices of $W$ - $W$-agents. Each $U$-agent  $u$ of $U$ has a linearly ordered preference list $L_u$ of a subset of $W$-agents $W'_u \subseteq W$ possibly containing ties and analogously each $W$-agent $w$ has a linearly ordered preference list $L_w$ of a subset of $U$-agents.
The majority of the terminology for stable matchings goes through for stable $b$-matchings. $v_1$ is acceptable for $v_2$ if $v_1$ is on $L_{v_2}$.
Instead of saying that some agent or vertex is single or free we will use the term {\bf unsaturated}: agent $v$ is unsaturated by a $b$-matching $M$ if $deg_M(v) < b(v)$.
A pair $(u,w)$ is blocking for a $b$-matching $M$ if (0) $u$ and $w$ are acceptable to each other and (1) $u$ is unsaturated or prefers $w$ to one of $W$-agents of $M(u)$ and if (2) $w$ is unsaturated or prefers $u$ to one of $U$-agents of $M(w)$. A $b$-matching $M$ is said to be stable if it does not admit a blocking pair.
As previously we are interested in finding a stable $b$-matching of largest size.
Let us also note that if for each $u$ in $U$ we have $b(u)=1$, then the problem is known under the name Hospitals-Residents problem or one-to-many stable matching problem.

An approximation algorithm for stable $b$-matchings is constructed analogously to the algorithm from the previous section.
$U$-agents play the role of men and $W$-agents play the role of women.
Each $U$-agent $u$ makes a proposal to each $W$-agent on $L_u$. Each $W$-agent $w$ stores information about $U$-agents currently matched with $w$ in a priority queue.
We translate the notions from the previous section to the current setting as follows.
If a saturated $U$-agent $u$ is matched with a $W$-agent $w$ and there is at least one unsaturated $W$-agent $w_1$
such that $w$ and $w_1$ are equally good for $u$, then $u$ and $w$ are said to be {\bf unstable} and $w_1$ is said to be {\bf
a satellite} of $u$. If $e=(u,w)$ is such a non-$M$-edge that $u,w$ are unsaturated  and there is at least one unsaturated $W$-agent $w_1$ such that $w$ and $w_1$ are equally good for $u$, then $e$ is called {\bf special}. 
We also have dangerous paths. Suppose we have a stable matching $M$, then  a path $P=(w,u_1,w_1,u)$ is called {\bf dangerous} if $(u_1,w_1)$ is in $M$, $(w,u_1),(w_1,u)$
are not in $M$, $w$ and $u$ are unsaturated, $u_1,w_1$ are saturated  and it is not true that
$u_1$ prefers $w_1$ to $w$ and $w_1$ prefers $u_1$ to $u$. Since $(w,u_1)$ is not blocking for $M$, $w$ is not better
for $u_1$ than any of the $W$-agents he is currently matched with and analogously $u$ is not better for $w_1$ than 
any of the $U$-agents he is currently matched with. Thus if $P$ is dangerous, then either $w,w_1$ are equally good for
$u_1$ and $(w,u_1)$ is called an {\bf equal} edge and $P$ a {\bf masculine dangerous path}, or $u, u_1$ are equally good
for $w_1$ and $(u,w_1)$ is called an equal edge and $P$ a {\bf feminine dangerous path}.
Analogously we define  {\bf blocking, unstable, f-equal} and {\bf bad} edges. 

Whenever $W$-agent $w$ receives a proposal from $u$, $w$ accepts $u$ if it is unsaturated or unstable (then a proper exchange also takes place) or compares $u$ to the worst $u'$ that is currently matched with $w$. Finding the worst $u'$
that is currently matched with $w$ takes $O(log(b(w)))$ time.

The running time of the algorithm is $O(m \log c)$, where $c=min\{max\{b(v): v\in U\},max\{b(v): v \in W\}\}$ and $m$ denotes the number of the edges and the approximation factor is $\frac{3}{2}$.

\noindent{\bf Acknowledgements} I would like to thank an anonymous referee for many helpful comments.


\vspace{-0.6cm}
\begin{thebibliography}{99}
\bibitem{GS} D.Gale, L.S.Shapley, College admissions and the stability of marriage,  American Mathematical Monthly, 69 (1962) 9-15.
\bibitem{Halld1} M.M.Halldorsson, R.W.Irving, K.Iwama, D.Manlove, S.Miyazaki, Y.Morita, S.Scott, Approximability results for stable marriage problems with ties, Theor. Comput. Sci. 306(1-3)(2003) 431-447.
\bibitem{Halld2} M.M.Halldorsson, K.Iwama, S.Miyazaki, H.Yanagisawa, Improved approximation results for the stable marriage problem, ACM Transactions on Algorithms 3(3) (2007).
\bibitem{Iwama1} K.Iwama, S.Miyazaki, K.Okamoto, A $(2-c \frac{\log n}{n})$-Approximation Algorithm for the Stable Marriage Problem, in: T.Hagerup, J.Katajainen (Eds.), Algorithm Theory - SWAT 2004, Humlebaek, 2004, pp. 349-361.
\bibitem{Iwama2} K.Iwama, S.Miyazaki, N.Yamauchi, A $(2-c\frac{1}{\sqrt(n)})$-Approximation Algorithm for the Stable Marriage Problem, in:  X.Deng, D.Du (Eds.): Algorithms and Computation, ISAAC 2005, Sanya, 2005, pp. 902-914.
\bibitem{Iwama3} K.Iwama, S.Miyazaki, N.Yamauchi, A $1.875$ - approximation algorithm for the stable marriage problem, in: N.Bansal, K.Pruhs, C.Stein (Eds.): Proceedings of the Eighteenth Annual ACM-SIAM Symposium on Discrete Algorithms, SODA 2007, New Orleans, 2007. SIAM 2007, pp. 288-297.
\bibitem{Kir} Z.Kiraly, Better and Simpler Approximation Algorithms for the Stable Marriage Problem, in:
D.Halperin, K.Mehlhorn (Eds.): Algorithms - ESA 2008,  Karlsruhe, 2008, pp. 623-634.
\bibitem{Lov} L.Lovasz, M.D.Plummer, Matching Theory, Ann. Discrete Math. 29, North-Holland, Amsterdam, 1986.
\bibitem{Manlove} D.Manlove, R.W.Irving, K.Iwama, S.Miyazaki, Y.Morita, Hard variants of stable marriage. Theor. Comput. Sci. 276(1-2)(2002) 261-279.
\bibitem{McDermid} E.McDermid: A 3/2-Approximation Algorithm for General Stable Marriage, in: S.Albers, A. Marchetti-Spaccamela, Y.Matias, S.E.Nikoletseas, W.Thomas (Eds.): Automata, Languages and Programming, 36th International Colloquium, ICALP 2009, Rhodes, Greece, 2009, pp. 689-700.
\bibitem{Y} H.Yanagisawa, Approximation algorithms for stable marriage problems, PhD thesis, Kyoto University, Graduate School of Informatics, 2007.
\end{thebibliography}
\end{document}